\documentclass[
amsmath,amssymb,aps,pre,twocolumn]{revtex4-2} 
\usepackage{graphicx}
\usepackage{float}
\usepackage{dcolumn}
\usepackage{bm}
\usepackage{hyperref}
\usepackage{xcolor}
\usepackage[mathlines]{lineno}
\usepackage[normalem]{ulem}

\begin{document}

\title{Polymorphism and crystal suppression\\in symmetric mixtures with entropic interactions}
\author{Santi Prestipino}
\affiliation{Dipartimento di Scienze Matematiche e Informatiche, Scienze Fisiche e Scienze della Terra,
Universit\`a degli Studi di Messina, viale F. Stagno d’Alcontres 31, I-98166 Messina, Italy}
\affiliation{Institute for Chemical-Physical Processes, National Research Council, Viale F. Stagno d’Alcontres 37, Messina 98158, Italy}
\author{Giuseppe Pellicane}
\affiliation{Dipartimento di Scienze Biomediche, Odontoiatriche e delle Immagini Morfologiche e Funzionali, Universit\`a degli Studi di Messina, I-98125, Messina, Italy}
\affiliation{School of Chemistry and Physics, University of Kwazulu-Natal, 3209 Pietermaritzburg, South Africa}
\affiliation{National Institute of Theoretical and Computational Sciences (NIThECS), 3209 Pietermaritzburg, South Africa}
\author{Enrique Lomba}
\affiliation{Instituto de Química Física Blas Cabrera, CSIC, Madrid, Spain}
\affiliation{NAFOMAT group, Universidade de Santiago de Compostela, Santiago de Compostela, Spain}
\author{Franz Saija}
\affiliation{Institute for Chemical-Physical Processes, National Research Council, Viale F. Stagno d’Alcontres 37, Messina 98158, Italy}
\date{\today}
\begin{abstract}
We investigate the phases of a binary mixture of hard spheres with equal diameter and negative non-additivity, favoring hetero-coordination in dense samples.
By performing extensive swap Monte Carlo simulations, we show that in an interval of small non-additivities, close to the eutectic point, the mixture fails to crystallize and the compressed fluid gets stuck in an amorphous state.
The onset of glass at intermediate compositions is associated with the competition between FCC and B2 (CsCl-type) crystal structures.
As the propensity of particles to hetero-coordinate grows further, however, the B2 crystal gives way to a B1 (NaCl-type) crystal at equimolarity, and again the mixture orders for all compositions.
We argue that this happens because of the underlying FCC symmetry of the B1 crystal.
These results, pertaining to a model with excluded-volume interactions only, point to the key role of entropy in generating amorphous structures in multi-component metallic, inorganic, and soft-matter glass formers.

\end{abstract}

\maketitle

\section{Introduction}
Since fabrication processes on the nano- and micro-scales are prohibitely costly, several materials of technological interest are more conveniently prepared by allowing two or more chemical species to self-assemble into crystalline or amorphous solids.
For example, mixtures of colloidal particles can form either photonic crystals or amorphous structures, both finding application in optical devices~\cite{yadav:2025,hu:2021}.
A common strategy to inhibit crystallization of colloids is to employ two particle species with different sizes.
Indeed, particles with two different diameters disrupt the regular packing needed for crystallization, enhancing the stability of the supercooled liquid and leading to glass formation.
Another class of amorphous materials are metal alloys, which can exhibit remarkable mechanical strength and resistance to corrosion, despite having molding features similar to plastics~\cite{schroers:2013,zhong:2014}.

There is a vast experimental literature showing that the ideal composition where to attain the amorphous state is the eutectic point~\cite{suryanarayana:2011,molinero:2006,kobayashi:2011,nandi:2016}.
Turnbull~\cite{turnbull:1969} made the conjecture that the glass transition temperature of a binary mixture scales with cohesive energy, implying that the ideal conditions to prevent the formation of an extended crystal upon cooling are met near the eutectic.
Kob and Andersen (KA) developed a popular microscopic model of binary mixture, where particle sizes and energy parameters are adjusted so as to inhibit crystallization near the eutectic of Ni-P~\cite{kob:1994,pedersen:2018,fomin:2024}.
Following the KA model, several other model mixtures have been proposed~\cite{toxvaerd:2009,valdes:2009,banerjee:2013,zhang:2015,ingebrigtsen:2021}, with the idea that the knowledge of the eutectic structure provides a theoretical framework for the development of new devices.
In these models, particles of different species attract each other, giving rise to a negative enthalpy of mixing, while unequal particle sizes cause packing frustration~\cite{toxvaerd:2009,banerjee:2013, zhang:2015, ingebrigtsen:2021}.

On the other hand, a central paradigm of liquid state theory is that the behavior of dense samples is largely determined by steric repulsion~\cite{hansen:2013}.
Indeed, the structure of most atomic crystals, as well as the essential phenomenology of glass formation, are captured by systems of impenetrable (hard) spheres.
Since there is no inherent energy involved in purely hard-sphere interactions, the equilibrium state at fixed density is determined by maximization of entropy:
under slow compression, the fluid inevitably crystallizes, as the gain in each particle's vibrational entropy outweighs the entropic penalty of locking the positions into a crystal.
In a binary mixture of spheres with diameters $\sigma_1$ and $\sigma_2$, 
we can enhance cohesion between the species by leveraging on a 1-2 bond length smaller than $(\sigma_1+\sigma_2)/2$~\cite{zhang:2015}.
In this case, the (negative) non-additivity parameter $\Delta$ is defined in terms of the cross diameter $\sigma_{12}=(1+\Delta)(\sigma_1+\sigma_2)/2$ (in the KA model, the effective size ratio is 0.88 and $\Delta\approx-0.15$).
However, a strong non-additivity would destabilize the crystal, just like size asymmetry.
One might wonder if crystallization is at least restored in a non-additive mixture of equal-sized spheres, regardless of the value of $\Delta$.
This issue is not only important from a practical perspective, namely for the design of new functional materials, but also theoretical, since the symmetric hard-sphere mixture is a fundamental statistical model that can shed light on the role of entropy in the glass-forming ability of colloids and metal alloys~\cite{zhang:2014}.

In this paper, we show that excluded-volume effects driven by non-additivity are not only capable, in a symmetric mixture, to suppress crystallization in an interval of negative $\Delta$ values, but also sufficient (for larger $|\Delta|$) to drive freezing into a B1 (NaCl-type) crystal, thereby creating the conditions for the recovery of crystal order at all compositions.
We argue below that these features are actually related to each other.
From a broader statistical-mechanical perspective, our study demonstrates that a complex emergent behavior can even arise in a minimal interaction model.

The paper is organized as follows.
We first describe the model and the simulation method used for its study (Section II).
In Section III we present and comment our results.
The final Section IV is devoted to conclusions and outlook.

\section{Model and method}
We use isothermal-isobaric Monte Carlo (MC) simulations~\cite{frenkel:2002} with swapping moves to study a non-additive and symmetric (i.e., $\sigma_1=\sigma_2=1$) mixture of two hard-sphere species, 1 and 2, for $\Delta=-0.1,-0.2,-0.3$ and compositions $c=N_2/(N_1+N_2)$ from 0.1 to 0.5 (clearly, $c$ and $1-c$ hold identical information).

Our samples contain about two thousand particles in total, which is sufficient for our purposes.
At each state point we typically perform $10^6$ MC cycles, while computing averages over the second half of MC trajectory.
Approximately 20\% of the attempted MC moves are swapping moves, whereby the positions of two 1 and 2 particles are exchanged;
then, the move is accepted by the standard Metropolis rule.
It is well established that swapping moves considerably expedite relaxation of supercooled fluids to equilibrium, reaching, for some models, speed-ups larger than $10^{10}$~\cite{berthier:2016,ninarello:2017,berthier:2019}, i.e., compatible with the time scales of relaxation in experiments.

Our method consists in running simulations in sequence, starting from the fluid at low $P$ and from the solid(s) at high $P$.
The pressure is increased or decreased in small steps, until abrupt freezing or melting is observed.
Finally, phase transitions are identified as inflection points in the EOS, and independently confirmed by structural indicators and visual inspection.

\section{Results}
Monodisperse hard spheres ($\Delta=0$) crystallize into a FCC solid, and we have verified that the same occurs for $\Delta=-0.05$ at all compositions.
The situation changes when a greater degree of particle overlap is allowed.

\begin{figure*}
\centering
\includegraphics[width=0.8\textwidth]{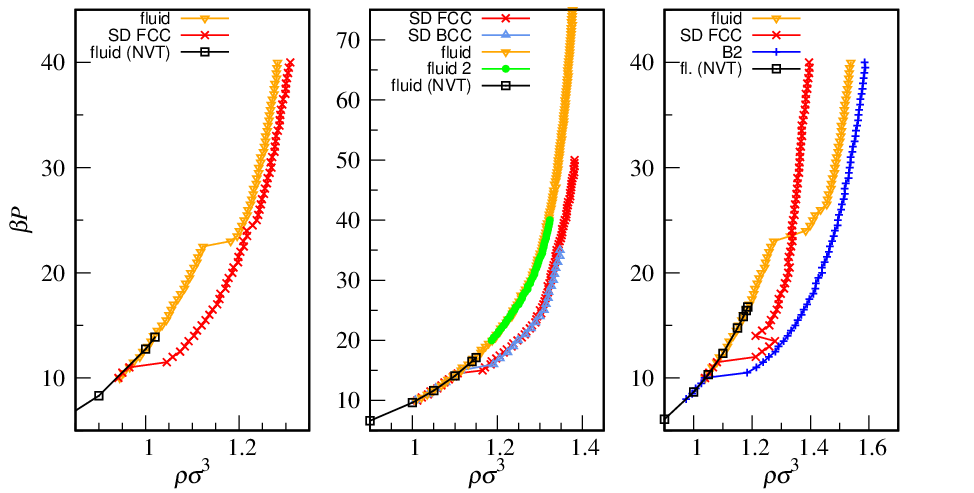}
\caption{\label{figsm1}
Mixture with $\Delta=-0.1$:
Fluid and solid branches of the EOS at $c=0.1$ (left), 0.3 (middle), and 0.5 (right).
The pressure $P$ is given in reduced, $k_{\rm B}T/\sigma^2=(\beta\sigma^2)^{-1}$ units ($k_{\rm B}$ is the Boltzmann constant, $T$ is the temperature, and $\sigma=1$ is the sphere diameter), and plotted as a function of the (reduced) total number density $\rho$.}
\end{figure*}

\begin{figure*}
\centering
\includegraphics[width=0.8\textwidth]{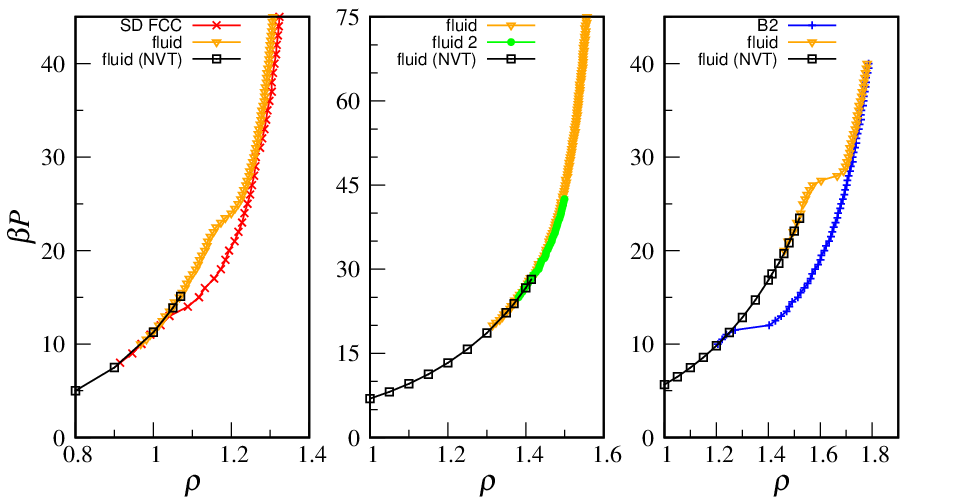}
\caption{\label{figsm2}
Mixture with $\Delta=-0.2$.
Fluid and solid branches of the EOS at $c=0.1$ (left), 0.3 (middle), and 0.5 (right).}
\end{figure*}

\begin{figure*}
\centering
\includegraphics[width=0.8\textwidth]{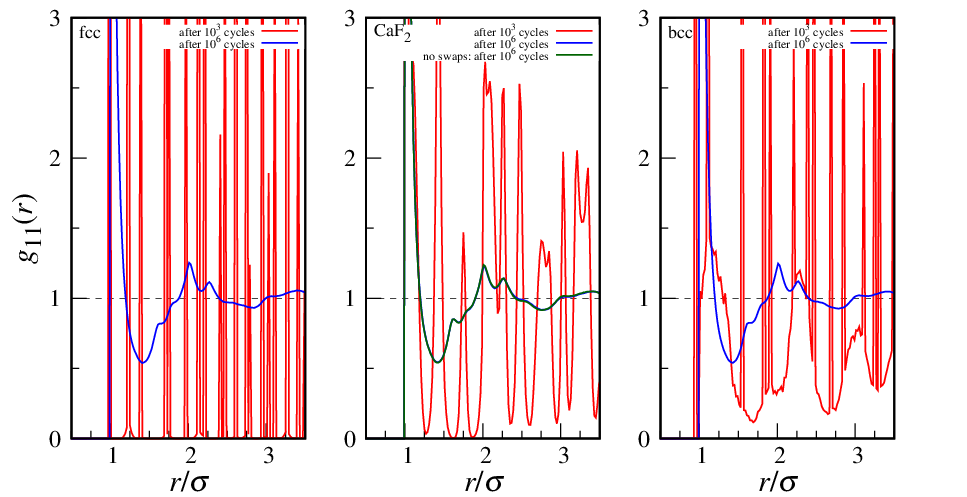}
\caption{\label{figsm3}
Mixture with $\Delta=-0.2$ at $c=0.3$.
Test of the mechanical stability of solids (left: SD FCC; middle: fluorite; right: SD BCC).
The RDF of the majority species is plotted after $10^3$ (red) and $10^6$ MC cycles (blue).}
\end{figure*}

\begin{figure*}
\centering
\includegraphics[width=0.8\textwidth]{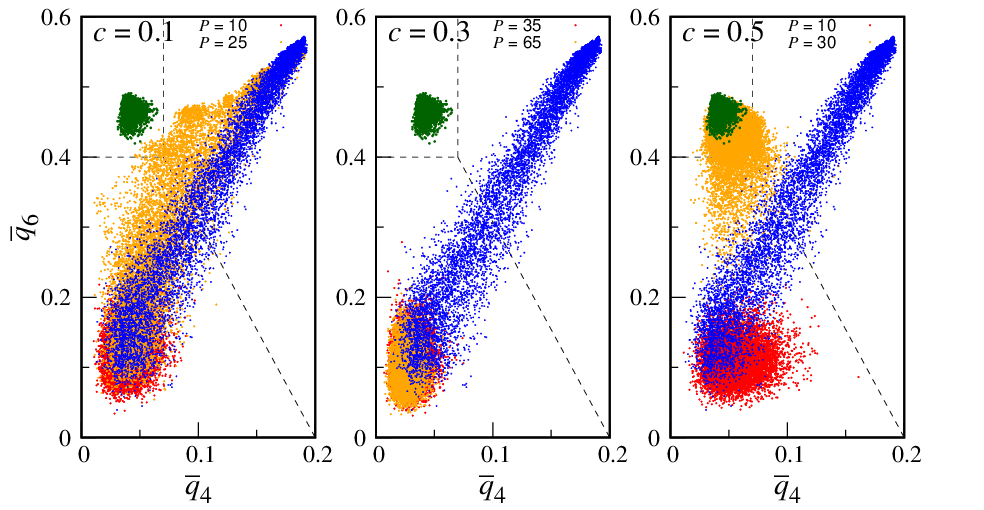}
\caption{\label{figsm4}
Mixture with $\Delta=-0.2$. Distribution of Lechner-Dellago parameters in various state points, for $c=0.1$ (left), 0.3 (middle), and 0.5 (right):
low- and high-pressure fluids (red and orange dots);
SD FCC at $c=0.1$ and $P=45$ (blue dots);
B2 crystal at $c=0.5$ and $P=45$ (green dots).}
\end{figure*}

\subsection{$\Delta=-0.1$}
To our knowledge, the only investigation of a symmetric hard-sphere mixture with negative non-additivity is the one reported in Ref.~\onlinecite{punnathanam:2006}, where the only case examined is $\Delta=-0.1$.
In that study, the mixture exhibits congruent freezing/melting (i.e., fluid and solid have identical compositions) and the eutectic composition is $c\simeq 0.3$.
The solids coexisting at the eutectic point are a substitutionally disordered (SD) FCC crystal and a SD BCC crystal, becoming a B2 (CsCl-type) crystal at $c=0.5$.
The B2 crystal has a BCC structure, with each species filling a different simple-cubic sublattice.
For $\Delta=0$, the BCC phase is loosely packed, whereas the density of a B2 mixture reaches the remarkable highest value of $2\sigma^{-3}$ if sufficient overlap is allowed between particles of different species.

In Fig.~\ref{figsm1} we plot the EOS of the mixture with $\Delta=-0.1$ for three compositions, 0.1 (left panel), 0.3 (middle panel), and 0.5 (right panel).
In agreement with Ref.~\onlinecite{punnathanam:2006}, we find that freezing occurs into a SD FCC solid at $c=0.1$ and into a B2 solid at $c=0.5$.
Looking at the left panel, we see that the phase transformation at $c=0.1$ occurs at different pressures along the fluid and the solid branch.
This effect (``hysteresis'') is due to the finite length of simulation runs, made worse by the relatively large pressure steps (0.5) employed.
However, our purpose here is not to locate the phase transition with extreme precision (a task that unavoidably requires performing ``exact'' free-energy calculations~\cite{frenkel:2002} or direct coexistence simulations~\cite{Smallenburg}).
Rather, our primary interest is identifying the structure of the incoming solid phase.
At $c=0.5$ (right panel) two different solid branches of the EOS are constructed (relative to SD FCC and B2), but it is clear from the path taken by the fluid branch under compression that the fluid transforms into a (defective) B2 crystal.
Nonetheless, the SD FCC crystal is mechanically stable; when depressurized, it eventually transforms into a (defective) B2 crystal before melting completely.
Instead, a B2 crystal with fewer or no defects melts at a slightly lower pressure.

Finally, for $c=0.3$ (middle panel) the fluid EOS is a smooth curve, suggesting the occurrence of a glass transition.
The disordered structure of the dense mixture at $c=0.3$ is demonstrated by the near-zero values of Steinhardt parameters~\cite{steinhardt:1983} and the fluid-like radial distribution function (RDF) of the majority species, $g_{11}(r)$.
We repeated the simulation by increasing the length of both relaxation and production runs by one order of magnitude, but the outcome stayed the same --- see the data points dubbed ``fluid 2'' in the legend.

While it was impossible to crystallize the mixture by compression at $c=0.3$, the SD FCC and SD BCC solids are mechanically stable at high density.
As reported in the middle panel of Fig.~\ref{figsm1}, the pressure ranges where the two crystals are (meta)stable are almost overlapping, which provides strong evidence of their proximity in free energy near the eutectic.
As a countercheck, a mixture with fluorite structure ($c=1/3$) is not mechanically stable.

\subsection{$\Delta=-0.2$}
By increasing hetero-coordination to $\Delta=-0.2$, we found it impossible to crystallize the mixture by compression in the whole range of compositions from 0.2 to 0.4.
Instead, the phase behavior at $c=0.1$ and 0.5 is identical to $\Delta=-0.1$.

The EOS of the mixture for the same three compositions above is shown in Fig.~\ref{figsm2}.
Overall, the behavior of the system parallels that for $\Delta=-0.1$:
the fluid mixture freezes into a SD FCC solid (respectively, a B2 solid) at $c=0.1$ (0.5), but fails to crystallize upon compression at $c=0.3$, even when the length of the runs is increased by one order of magnitude (``fluid 2'' data).
As anticipated, the main difference from $\Delta=-0.1$ is that simulating the solid phases for $c=0.3$ now proves impossible.
Specifically, we examined three cases: SD FCC, fluorite, and SD BCC.
After preparing the mixture as a perfect crystal at $P=40$, a million MC cycles later the system is completely melted (namely, the RDF has the same fluid-like shape in every case, see Fig.~\ref{figsm3}).

To support the conclusions drawn from the previous Figs.~\ref{figsm2} and ~\ref{figsm3}, we report in Fig.~\ref{figsm4} the values at regular intervals along the MC trajectory of the Lechner-Dellago (LD) parameters $\overline{q}_4$ and $\overline{q}_6$~\cite{lechner:2008}.
In simple terms, these quantities provide a local measure of crystalline order in the orientation of nearest-neighbor bonds throughout the system.
Upon averaging over particles and MC time, the joint distribution $P(\overline{q}_4,\overline{q}_6)$ of LD parameters bears a clear mark of the structure of the mixture:
as we see in Fig.~\ref{figsm4}, distinct bulk phases have distributions with disjoint support.
It is interesting to look at the shape of $P(\overline{q}_4,\overline{q}_6)$ as a function of pressure along the fluid branch.
At low pressure, the values of $\overline{q}_4$ and $\overline{q}_6$ are concentrated in the bottom-left corner of the $\overline{q}_4$-$\overline{q}_6$ plane (the mixture is fluid).
At high pressure, in those cases where the mixture freezes upon compression (left and right panels) the shape of $P(\overline{q}_4,\overline{q}_6)$ conforms to that of the bulk solid phase.
The residual differences between the two distributions are due to the defects present in the solid emerging from the fluid.
The case $c=0.3$ is different, since the shape of $P(\overline{q}_4,\overline{q}_6)$ remains fluid-like also at high pressure.

\begin{figure}
\centering
\includegraphics[width=0.48\textwidth]{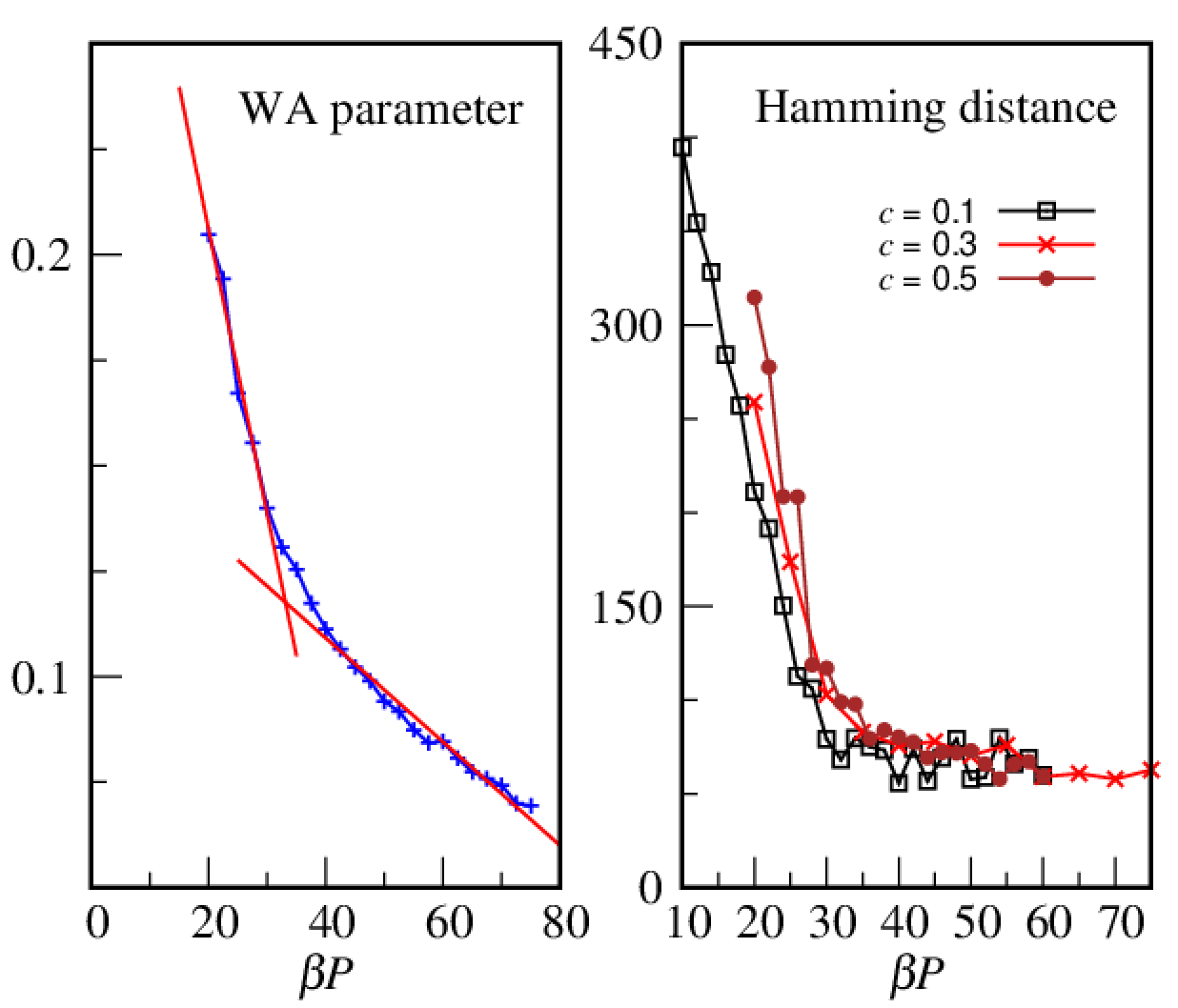}
\caption{\label{fig1}
Mixture with $\Delta=-0.2$ and $c=0.3$.
Left: WA parameter along the fluid branch.
Right: Hamming distance for various compositions, plotted as a function of $\beta P$.}
\end{figure}

To assess the existence of a glass transition for $c=0.3$, we plot in Fig.~\ref{fig1} the Wendt-Abraham (WA) parameter~\cite{wendt:1978} and the Hamming distance~\cite{rodríguez:2007} as a function of pressure.
The WA parameter, defined as the ratio between the first minimum and the first maximum of $g_{11}(r)$, exhibits a characteristic change when the system undergoes structural arrest.
Indeed, we see a kink in our data at $P\approx 35$, where the WA parameter ($\approx 0.12$) lies in the range expected for the glass transition~\cite{fan:2024,celtek:2021}.  
Similarly, the Hamming distance (a measure of dissimilarity between the initial and final configurations in the simulation run) flattens out beyond $P\approx 35$, settling at a level comparable to the solids at $c=0.1$ and 0.5.
At the same composition $c=0.3$ the behavior of the mixture for $\Delta=-0.1$ is qualitatively the same as for $\Delta=-0.2$.
We will cover the molecular-dynamics study of the glass transition of our mixture in a separate article.

\begin{figure*}
\centering
\includegraphics[width=0.65\textwidth]{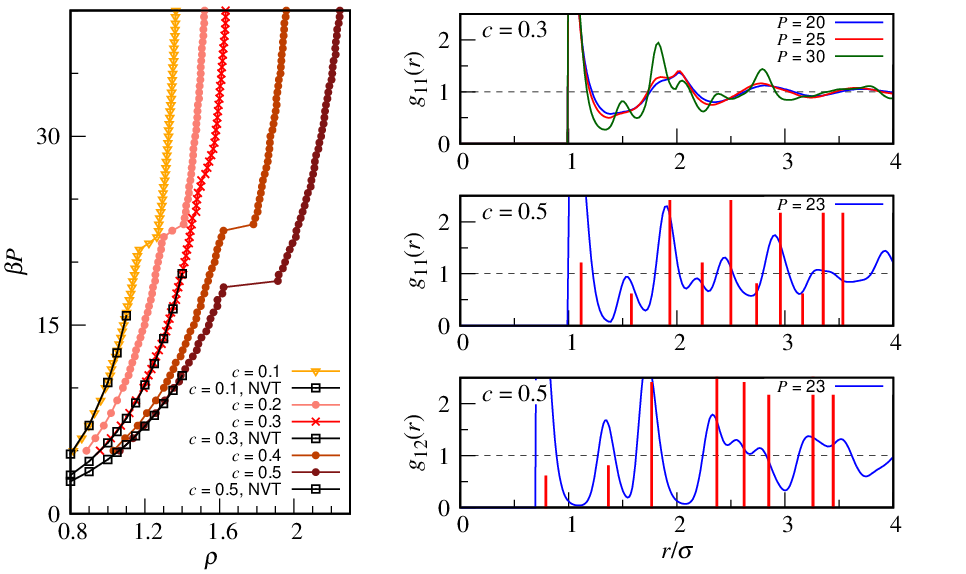}
\caption{\label{fig2}
Mixture with $\Delta=-0.3$.
Left: fluid EOS for $c=0.1,0.2,0.3,0.4,0.5$.
The inflection point on each curve marks the transition of the overcompressed fluid into a solid phase.
Top right: $g_{11}(r)$ for $c=0.3$ and $P=20,25,30$.
Middle right: $g_{11}(r)$ for $c=0.5$ and $P=23$.
The red lines mark the neighbor distances in a perfect FCC crystal of same density (their length is one tenth of the number of neighbors).
Bottom right: $g_{12}(r)$ for $c=0.5$ and $P=23$.
The red lines are now located at FCC distances as measured from an octahedral site.}
\end{figure*}

\subsection{$\Delta=-0.3$}
\begin{figure*}
\centering
\includegraphics[width=7cm]{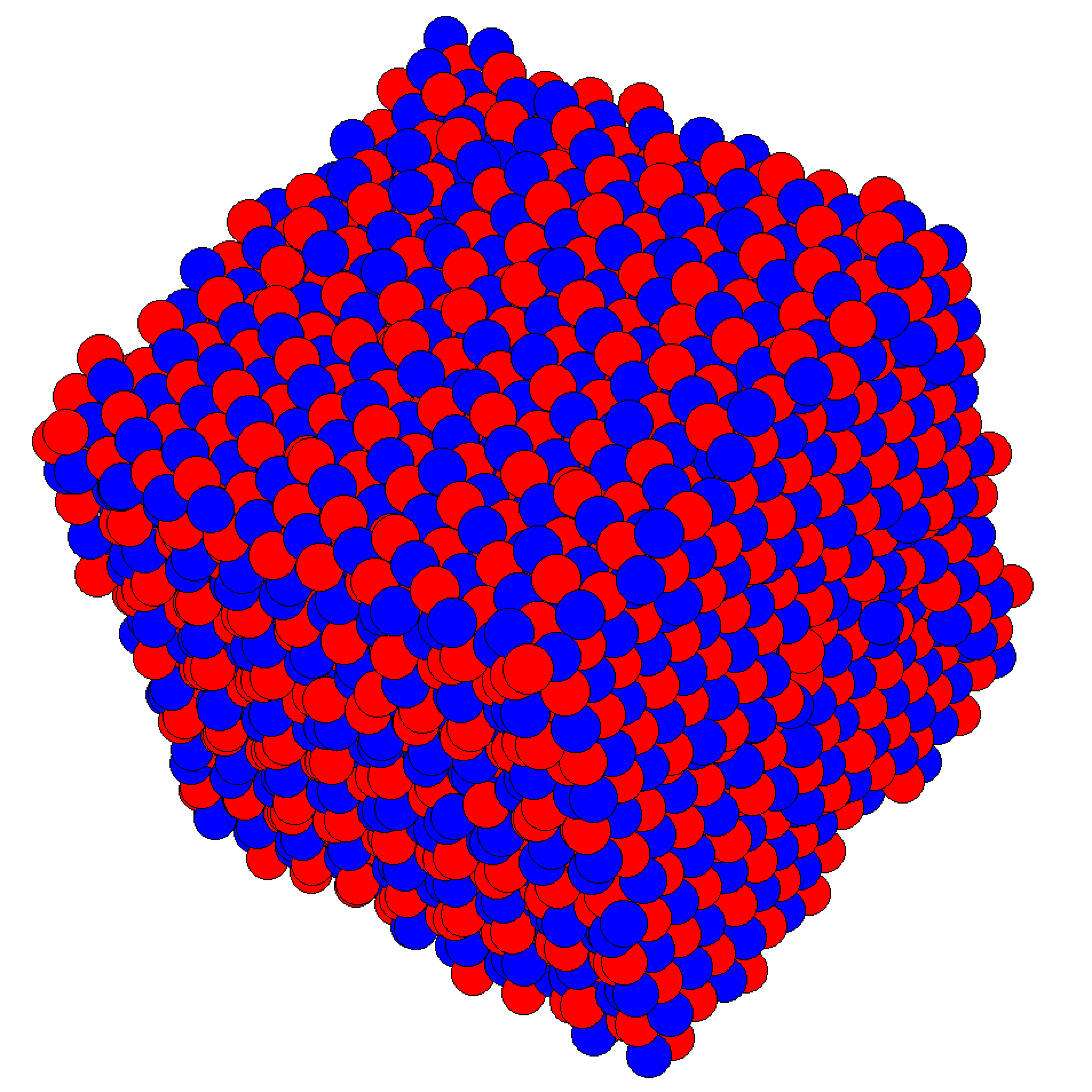}\qquad
\includegraphics[width=7cm]{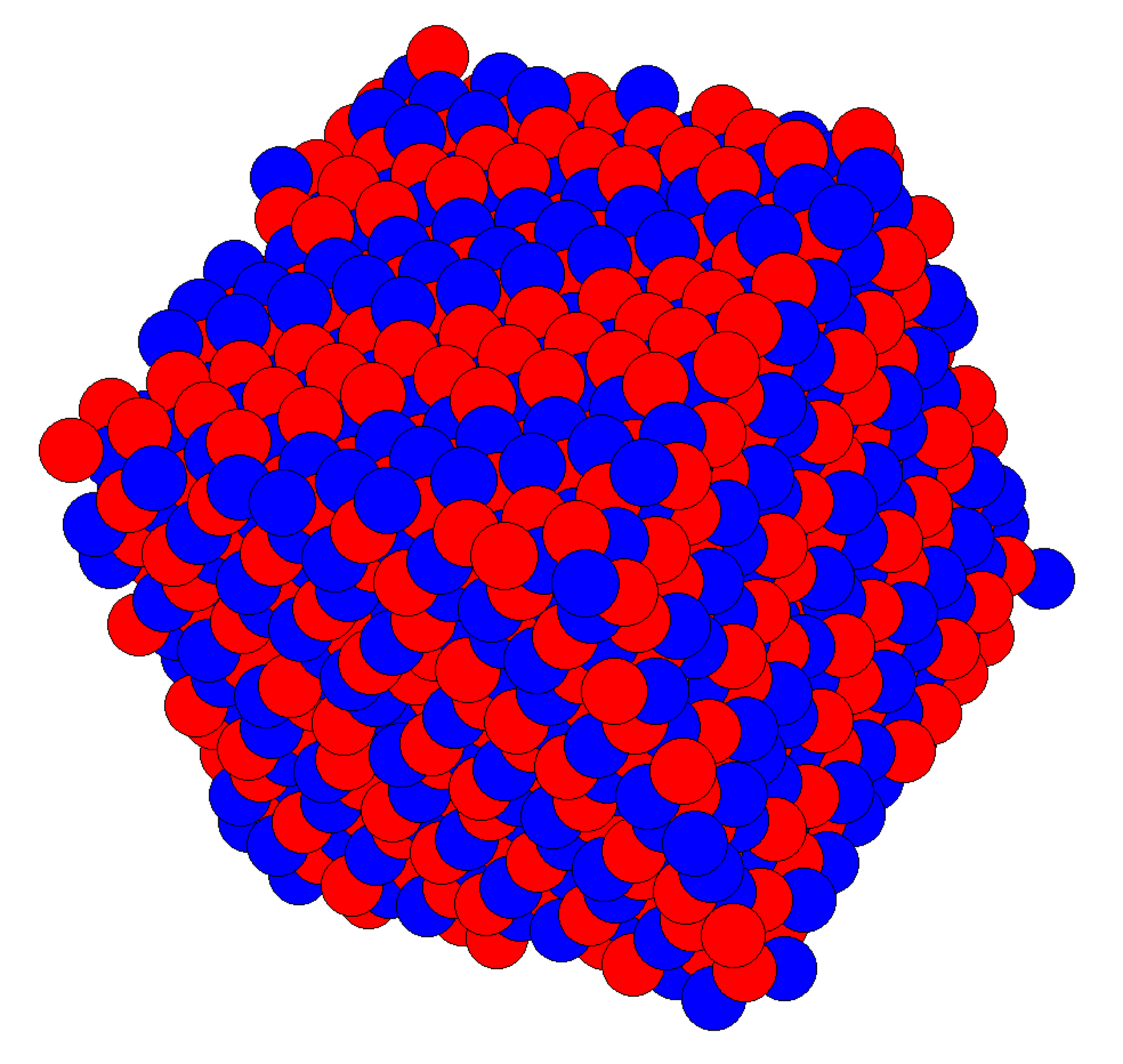}
\caption{\label{figsm5}
Snapshot of the compressed fluid mixture at $c=0.5$. Left: $\Delta=-0.2$ and $P=33$ (B2 crystal). Right: $\Delta=-0.3$ and $P=23$ (B1 crystal).}
\end{figure*}

When moving to $\Delta=-0.3$, the phase behavior of the mixture changes radically.
At $c=0.3$ the EOS develops an inflection point at $P\approx 27$ (see left panel of Fig.~\ref{fig2}) which heralds the transition to a solid phase, as witnessed by the sharp peaks of $g_{11}(r)$ (top right panel).
While at $c=0.1$ the mixture still freezes into a SD FCC crystal, at $c=0.5$ we see the onset of a new phase, characterized below as B1, consisting of two interpenetrating FCC crystals.
The striped pattern of this phase is evident in Fig.~\ref{figsm5} (right panel), where the two species are arranged on alternating planes running in the [111] direction;
within each plane, particles are triangularly close-packed, such as occurs in the (111) planes of a FCC crystal.

Similarly to B1, also a B2 crystal exhibits, in its simple-cubic sublattices, a triangular arrangement of like particles on (111) planes, with ABC stacking along [111].
However, on these planes particles are loosely packed.
In order to transform a B2 crystal into a B1 crystal we should ideally perform a compression on each (111) plane, followed by a suitable expansion along [111], as we shall see in more detail in the next section.
At the end of this process, each particle in one FCC sublattice will be trapped in the octahedral cavity formed by particles sitting on the other FCC sublattice.

When $\Delta<\sqrt{2}/2-1\simeq -0.293$ an equimolar mixture formed by two interpenetrating FCC lattices filled with particles of different species reaches its highest density of $2\sqrt{2}\sigma^{-3}\simeq 2.828\sigma^{-3}$ (see Section III.D).
When this happens, nearest-neighbor like particles are at contact.
In this solid, corresponding to a close-packed B1 crystal, triangularly-ordered lattice planes filled with particles of one species are alternated with planes filled with particles of the other species, with ABC stacking along [111].
$2\sqrt{2}\sigma^{-3}$ is much larger than the maximum density allowed to a B2 crystal; whence the easy prediction that, upon increasing $|\Delta|$, direct freezing into a B1 crystal becomes likely, due to the enthalpic advantage of B1 over B2.

\begin{figure*}
\centering
\includegraphics[width=0.65\textwidth]{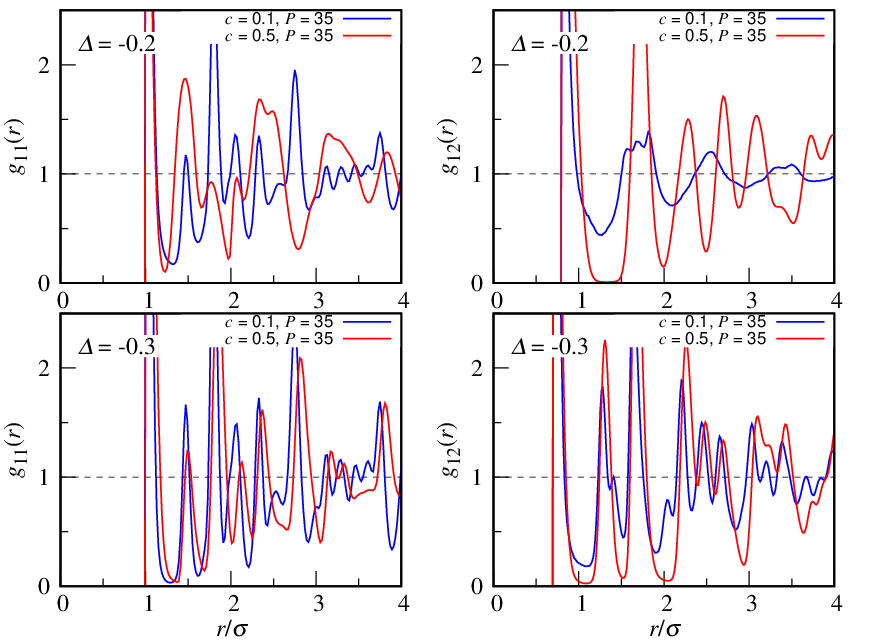}
\caption{\label{fig3}
Mixture with $\Delta=-0.2$ (top panels) and $-0.3$ (bottom panels), at $P=35$.
Left: $g_{11}(r)$ for $c=0.1$ and 0.5.
Right: $g_{12}(r)$ for $c=0.1$ and 0.5.}
\end{figure*}

To characterize in more detail the structure of the solid phase at $c=0.5$, in the lower right panels of Fig.~\ref{fig2} we plot the RDF of 1-1 and 1-2 pairs.
While the location of $g_{11}(r)$ peaks align with the characteristic distances of a FCC lattice, the peaks of $g_{12}(r)$ agree well with the same distances, but measured from an interstitial octahedral site.
This evidence uniquely identifies the new crystal as B1.

Now, we reconsider the behavior of the mixture at intermediate compositions.
We argue that crystal suppression for $\Delta=-0.2$ is due to the competition between FCC and BCC, pushing for incompatible short-range patterns in the mixture.
When this conflict is absent, as for $\Delta=-0.3$, crystallization occurs for all compositions.
To substantiate our belief, in Fig.~\ref{fig3} we plot a few RDFs of like and unlike particles at high pressure, for $\Delta=-0.2$ (top) and $\Delta=-0.3$ (bottom).
While the profiles at $c=0.1$ and $c=0.5$ are clearly incompatible for $\Delta=-0.2$, they are commensurate with each other for $\Delta=-0.3$.
This provides a rationale for the high-pressure structure at intermediate compositions, which is disordered for $\Delta=-0.2$ due to the irresolvable frustration between the FCC and BCC crystal symmetries.

Even though not realistic, we briefly discuss what happens when $|\Delta|$ is bigger than 0.3.
Eventually, there comes a point where a particle of species 1 can go through the center of a triangle formed by particles of species 2 in contact with each other, and vice versa.
When this happens, the FCC crystals of the two species cease to be interlocked and can freely diffuse one through the other, as we indeed observed for $\Delta=-0.5$.

\subsection{The reverse B\"urger’s distortion}

\begin{figure}[t]
\centering
\includegraphics[width=6 cm]{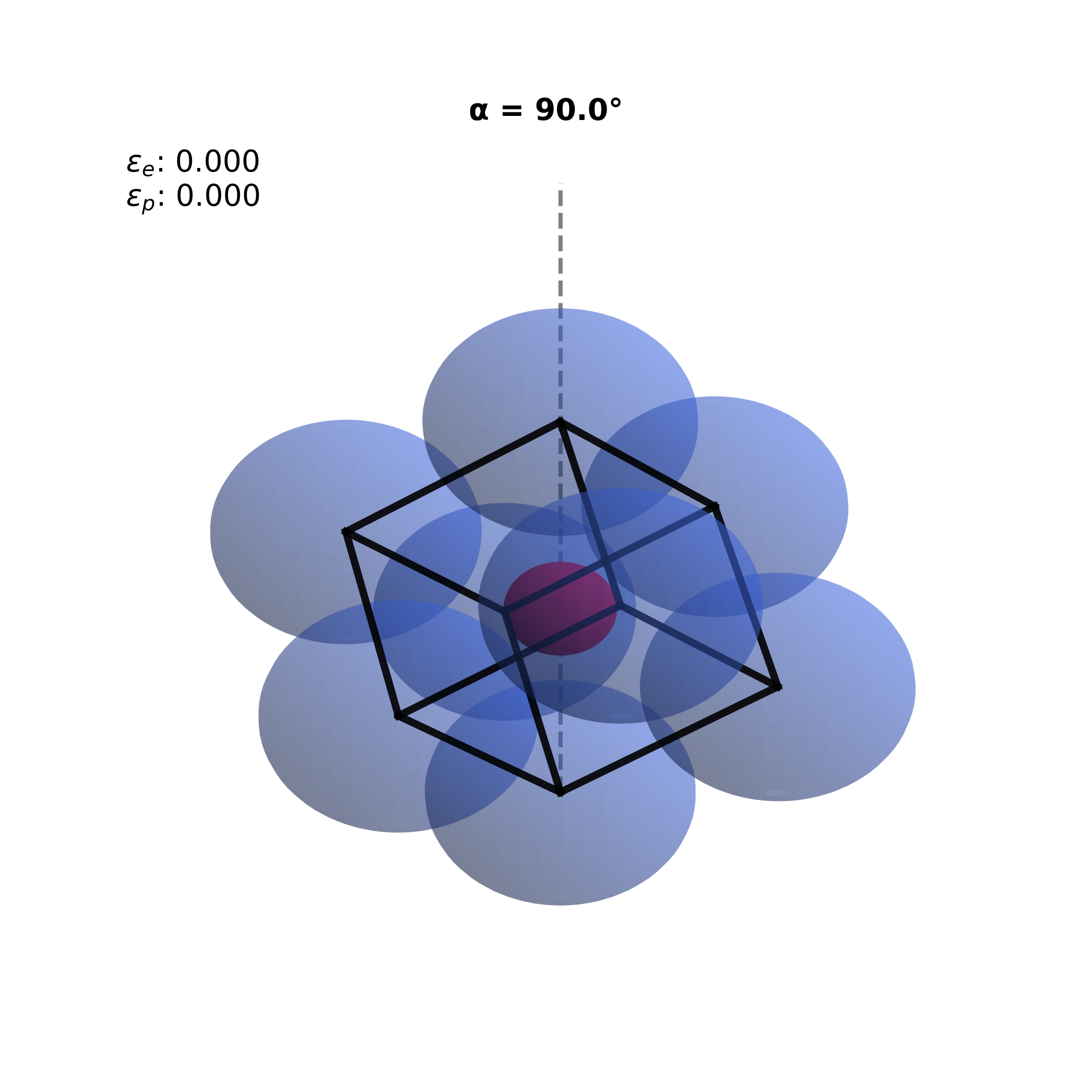}
\includegraphics[width=6 cm]{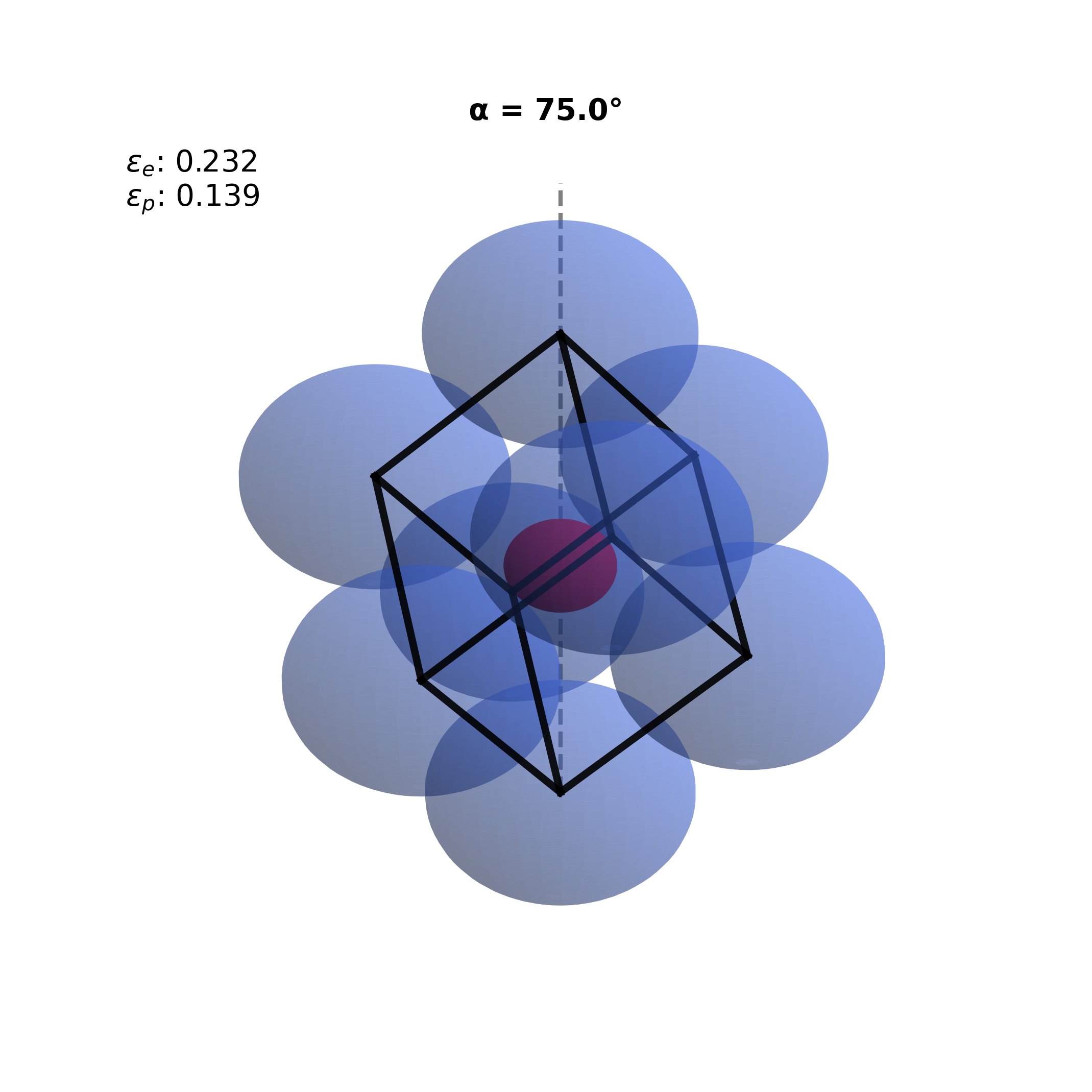}
\includegraphics[width=6 cm]{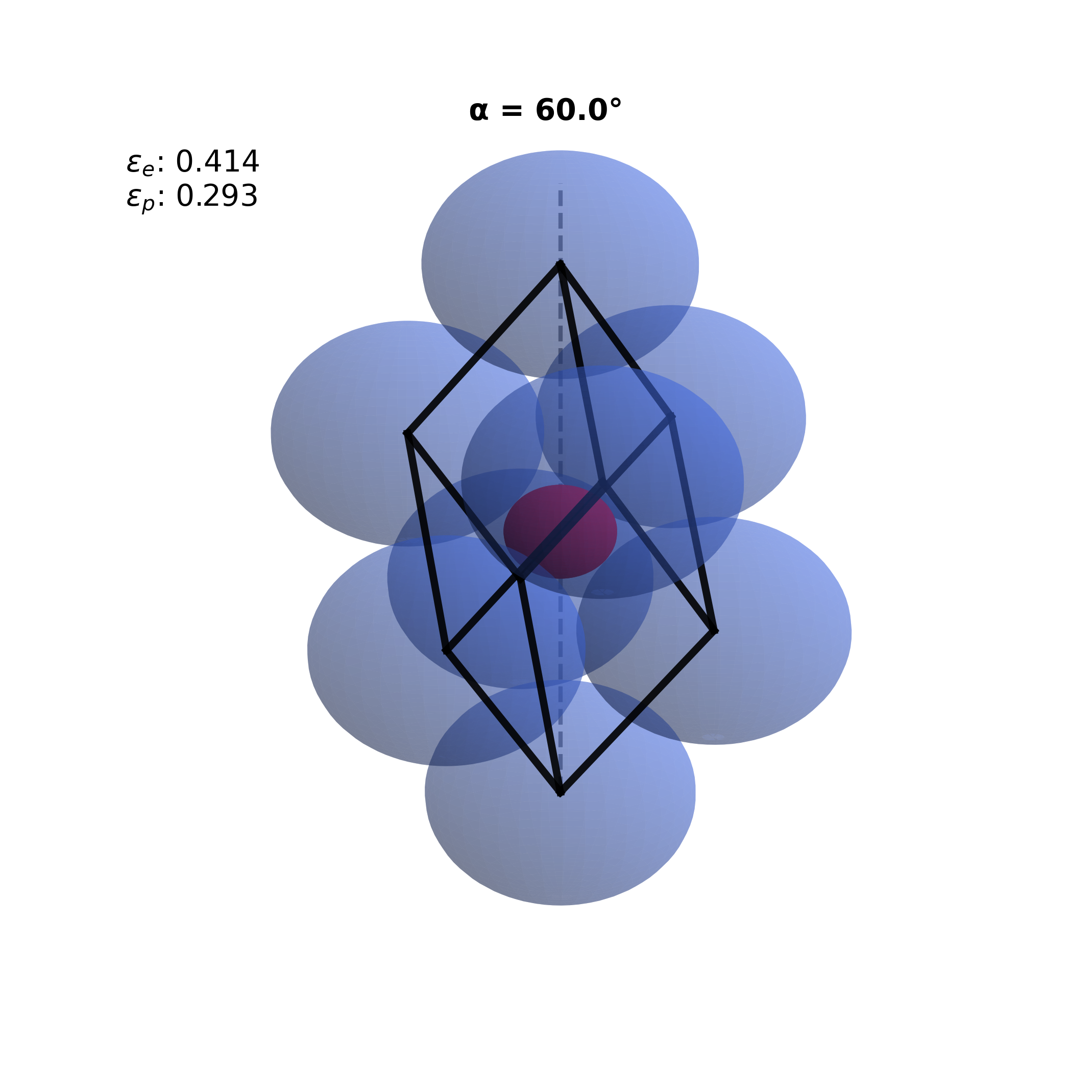}
\caption{Fixed-$a_R$ distorsion:
rhombohedral cell for angles $\alpha=90^\circ$ (top), $75^\circ$ (middle), $90^\circ$ (bottom) with particles of the same species at fixed distance $\sigma$, and $\Delta=-0.293$.
See supplementary material for the choice of the planar and axial expansions, $\epsilon_{\rm p}$ and $\epsilon_{\rm e}$~\cite{supplementary}.} 
\label{figg12}
\end{figure}

A more physically sound approach to understand the onset of B1 from B2 is to fix the distance between two nearest-neighbor particles of same species, which amounts to keep the lattice parameter $a_R$ of the rhombohedral cell fixed, as the rhombohedral angle $\alpha$ changes, through a reverse B\"urger’s distorsion \cite{buerger:1951}, from $90^\circ$ (B2 lattice) to $60^\circ$ (B1 lattice).
During this transformation, the crystal expands along the $[111]$ direction and is isotropically compressed in the $(111)$ plane, as reported in Fig.~\ref{figg12}.
This scenario reminds of a displacive martensitic transition, where the energy penalty (ruled by the bulk modulus) for stretching or compressing a bond is much larger than the penalty (ruled by the shear modulus) for bending a bond~\cite{supplementary}.

Hereafter we refer to the two species of particles as cations and anions, in analogy with the ionic compounds crystallizing into the B1 (Rocksalt) and B2 (Cesium Chloride) lattices.
We apply the reverse B{\"u}rger’s distortion to our model~\cite{supplementary}, in which the distance of minimum approach between two particles of unlike species is $(1+\Delta)\sigma$, while being $\sigma$ for particles of the same species.
If the nonadditive parameter $\Delta$ is not negative enough, there is no sufficient overlap between two unlike particles, and the equilibrium structure at equimolarity is the BCC lattice, so that each particle has eight nearest-neighbor particles of the other species:
this is the B2 structure.
However, if $\Delta$ becomes negative enough, it is possible for the BCC lattice to deform so that it splits into two interpenetrating FCC lattices, selectively occupied by particles of the two species, thus making the solid phase equivalent to the one of NaCl (B1 structure).

To understand the stability of the B2 versus B1 phase, we assume for clarity that in the crystal phase particles belonging to the same species keep their distance fixed at exactly its close-packed limit ($\sigma$).
Since fixing the distance can be viewed as fixing the bond distance between two like particles, we refer to this reverse B{\"u}rger’s distortion as an isodesmic pathway --- which amounts to hold the rhombohedral lattice parameter $a_R$ fixed, so as to create a perfect ``hinge'' model where the cation-cation distances along the edges of the rhombohedron are forced to remain constant, while the angles between them pivot from 90$^\circ$ to 60$^\circ$~\cite{supplementary} (see Fig.~\ref{figg12}).
Actually, it is shown in the supplementary material~\cite{supplementary} that the analytic expression for the cation-cation distance is entirely independent of the reaction coordinate $\alpha$, and equal to the close-contact distance $\sigma$ between two like particles.
This implies that the reverse B{\"u}rger transformation at fixed $a_R$ is a constant-bond-length transformation for the same-species sublattice (namely, an isodesmic pathway).
In other words, during this transformation the atoms slide past each other, changing the shape and volume of the crystal, while the distance between two nearest cations or anions remains rigidly locked at $\sigma$.

To explore the reverse B{\"u}rger mechanism (from B2 to B1) under these rules, we map the exact geometric limits of the rhombohedral cell as it shrinks in the $(111)$ plane and expands along $[111]$.
For arbitrary $\Delta$, the constraint for the distance $d_{AA}$ ($d_{AB}$) between two like (unlike) particles is $d_{AA}\ge \sigma$ ($d_{AB}\ge(1+\Delta)\sigma$).
In the supplementary material~\cite{supplementary} we show that during the B{\"u}rger distortion we can track any state of the rhombohedral unit cell using variables $L$ (the distance between two adjacent $\{111\}$ layers filled with particles of either species) and $H$ (the vertical height of the three-layer repeating unit along $[111]$). 
To perform a ``close-packed'' reverse transformation, we must continuously shrink the planar distance $L$ (from B2 to B1), while adjusting the height $H$, so that we never violate the minimum-distance constraint.
We find that the latter condition is met all along the fixed-$a_R$ transformation by simply requiring that the distance $(1+\Delta)\sigma$ between two unlike spheres never overcomes $\sqrt{2}\sigma/2$:
\begin{equation}
(1+\Delta)\sigma\le\frac{\sqrt{2}}{2}\sigma\implies\Delta\le-0.293.
\label{eq.18}
\end{equation}
In Fig.~\ref{figg12}, we show the rhombohedral cell for angles $\alpha=90^\circ$ (B2 phase), $75^\circ$ (intermediate step), and $60^\circ$ (B1 phase), representing the lattice distorsion with particles of the same species at fixed distance $\sigma$, and $\Delta=-0.293$. The red sphere has a diameter $\sigma_B=0.414\sigma$, since $(1+\Delta)\sigma=(\sigma+\sigma_B)/2=0.707\sigma$, which exactly represents the maximum diameter of a sphere inside the octahedral cavity of a FCC lattice ($0.414\simeq\sqrt{2}-1$).
Thus, when $\alpha=60^\circ$ the blue spheres are still in close contact and the red sphere touches the six equatorial blue spheres forming the octahedron (see Fig.~\ref{figg12}).

If $\Delta\le-0.293$, the constant-$a_R$ path perfectly bridges the B2 and B1 phases without requiring any transient over-expansion of the same-species sublattice.
Indeed, in that case the minimum distance between two unlike spheres is small enough to allow for the transformation to proceed without a fault along the fixed-$a_R$ path, landing perfectly in the B1 state, because the interstitial atoms exactly fit the voids.
Otherwise, as $L$ approaches $\sigma$ the two species will penetrate into each other, and to bring the transformation forward the crystal would be forced to stretch the same-species distance, by elongating $H$ beyond its natural boundary.

To sum up, the constant-$a_R$ reverse B\"urger’s distortion is the pathway through which the solid mixture lowers its Gibbs free energy by reducing the volume.
This condition is first met at $\Delta=-0.293$, as shown by Eq. \ref{eq.18}.
At $\Delta=-0.3$, the B2 crystal is simply too loose-packed and the enthalpic cost for all that empty space too high.
The reverse B\"urger’s distortion allows the mixture to get rid of that wasted space:
starting from the B2 phase, the atoms slide past each other, changing the shape and volume of the cell at the expenses of the central atom's freedom;
the originally symmetric 8-atom cage is distorted, as two neighbors are pushed away along [111] and the six neighbors on the (111) plane squeeze inward toward the central atom, which eventually becomes trapped in the octahedral cage of its unlike neighbors~\cite{supplementary}.
Conversely, when $\Delta\approx -0.2$ the B2 phase is sufficiently compact, while at the same time providing particles with enough room to wiggle.

\section{Conclusions}
We have shown that changes in a single parameter $\Delta$, measuring the affinity between unlike particles, can generate a wealth of structures in a symmetric mixture of hard spheres.
As $|\Delta|$ grows, the highest number of type-1 spheres that tightly overlap with a central type-2 sphere decreases, and this in turn establishes (in concert with composition) the kind of aggregates arising locally in the dense mixture.
In particular, a glass is formed when the most compact aggregates cannot fill space in a regular fashion.
The values of $\Delta$ (between $-0.2$ and $-0.1$) at which a glass transition occurs in our model are common in metal alloys~\cite{zhang:2015}.
Our study indicates that an important drive toward a glassy state in these materials could be steric frustration combined with entropy maximization (i.e., the organizing principle at play in hard-particle systems), which may be regarded as a third mechanism besides size disparity and negative enthalpy of mixing.
We thus provide a theoretical basis for recent experimental work~\cite{zhang:2026}, identifying a possible origin of vitrification in the competition between incommensurate crystal structures near the eutectic point.
When there is no frustration, as in the case of the mixture with $\Delta=-0.3$ (which features a B1 crystal at equimolarity), the glass transition simply disappears.

Our predictions can be tested in real materials.
For example, a strong inter-species attraction can be obtained by grafting complementary DNA strands onto colloidal particles~\cite{matos:2023}.
Another possibility is to implant either of two polymer brushes on the surface of colloidal particles, choosing polymers with a strong mutual affinity.

An unexpected finding of our study is the pressure-driven solidification of the equimolar mixture with $\Delta=-0.3$ into a B1 crystal.
This phase exhibits a highly symmetric and compact structure, along with a complete optical band gap that is beneficial for photonic applications~\cite{joannopoulos:2008}.
Based on our results, the synthesis of the B1 phase will not require the delicate atomic weaving needed by the more complex (and desirable for applications) MgCu$_2$ Laves phase --- where the two species are arranged in a diamond lattice and its gaps, which is hard to grow precisely due to the risk of kinetic trapping.

\begin{acknowledgments}
This work has benefited from the use of the computers made available by the PO-FESR 2007-2013 Project MedNETNA (Mediterranean Network for Emerging Nanomaterials).
E.L. acknowledges financial support from Grant No. PID2023-151751NB-I00, funded by MICIU/AEI/10.13039/501100011033 and by ``ERDF A way of making Europe'', and Galicia's Supercomputing Center (CESGA) for the generous access to their computer facilities.
\end{acknowledgments}

\end{document}